# Plasmonic Anti-Hermitian Coupling for Controlling Light at the Nanoscale

Shuang Zhang[1, 2*], Ziliang Ye[1*], Yuang Wang[1*], Yongshik-Park[1], Guy Bartal[1], Michael Mrejen[1], Xiaobo Yin[3] & Xiang Zhang[1, 3]⸸

1. 5130 Etcheverry Hall, Nanoscale Science and Engineering Center, University of California, Berkeley, California 94720-1740, USA

2. School of Physics and Astronomy, University of Birmingham, Birmingham, B15 2TT, UK

3. Materials Sciences Division, Lawrence Berkeley National Laboratory, 1 Cyclotron Road Berkeley, California 94720

**Open quantum systems consisting of coupled bound and continuum states have been studied in a variety of physical systems, particularly within the scope of nuclear, atomic and molecular physics. In the open systems, the effects of the continuum decay channels are accounted for by indirect non-Hermitian couplings among the quasi-bound states. Here we explore anti-Hermitian coupling in a plasmonic system for spatially manipulating light on the nano-scale. We show that by utilizing the anti-Hermitian coupling, plasmonic antennas closely packed within only λ/15 separations can be individually excited from the far field, which are otherwise indistinguishable from each other.**

⸸ To whom correspondence should be addressed. E-mail: xiang@berkeley.edu

* These authors contributed equally to this work

The indirect coupling among quasi-bound states through common continuum decay channels have been widely studied in various open quantum systems, in particular for the investigation of the resonance phenomena in nuclei, atoms, molecules, and quantum dots [1-3]. Interesting features of these systems have been observed, such as the restructuring of the eigen-states with contrasting lifetimes in the system: some long-lived states are deprived of the coupling strength to the decay channels, while others with enhanced coupling to the decay channels having very short lifetime, i.e. the so called "super-radiant" states that was first proposed by Dicke for describing the coherent spontaneous radiation of a gas confined to a sub-wavelength scale [4]. Spectroscopically, this phenomenon is manifested as sharp resonances super-imposed on the broad super-radiant states, as observed, for example, in the decay of compound nuclear states in certain nuclear reactions [5-7].

A general treatment of the open systems, the standard projection formulism, divides the Hilbert space of an open system into two subspaces, $\{Q\}$ subspace that consists of interior quasi-bound (discrete) states $|q\rangle$ and $\{P\}$ subspace that only consists of the continuum decay channels $|c\rangle$, with $Q$ and $P$ being the projection operators of the two subspaces [8]. The effective Hamiltonian acting on the bound states by taking into account the indirect coupling mediated by the continuum decay channels can be written as [9],

$$H_{eff} = H_{QQ} + H_{QP}(E - H_{PP})^{-1} H_{PQ} \qquad (1)$$

where $H_{QQ}$ is the Hamiltonian projected in the subspace of quasi-bound states, $H_{QP}$ and $H_{PQ}$ are the coupling matrices between the $Q$ and $P$ subspaces, and $(E - H_{PP})^{-1}$ is the propagator in the subspace of open channels. The second term on the RHS of Eqn (1) represents the indirect interaction among the bound states mediated by the open (continuum) channels. The matrix elements of the effective Hamiltonian can be further expressed as [1],

$$\langle q_1 | H_{eff} | q_2 \rangle = \langle q_1 | H_{QQ} | q_2 \rangle + \sum_{c=1}^{K} P.V. \int dE' \frac{V_{q_1}^c V_{q_2}^c}{E - E'} - \frac{i}{2} \sum_{c=1}^{K} V_{q_1}^c V_{q_2}^c$$

or in the matrix form,

$$H_{eff} = H_{QQ} + P.V. \int dE \frac{VV^+}{E - E'} - \frac{i}{2} VV^+ \qquad (2)$$

where P.V. denotes the Cauchy principal value of the integral, $V_q^c = \sqrt{2\pi}\langle q|H_{QP}|c\rangle$, $K$ is the number of decay channels. Thus, within the quasi-bound states subspace, the effective Hamiltonian of the system contains an imaginary, or anti-Hermitian coupling matrix ($-iVV^+/2$) that arises from the indirect coupling among the bound states mediated by the open channels [10,11]. For an open quantum system consisting of $M$ bound states coupled to $K$ continuum decay channels, the coupling matrix $V$ between the discrete and continuum subspaces is of dimension $M \times K$, and the indirect anti-Hermitian coupling matrix $-iVV^+/2$ has a dimension of $M \times M$, but with a rank of $K$ in the case of $M > K$. For an open system with a single common decay channel ($K = 1$), $V = [V_1, V_2, ..V_p..]^T$ is simply a column vector. Note that the Cauchy principal value of the indirection coupling (2nd term on RHs of Eqn. 2), which is real and generally non-zero, leads to a shift of the energies of eigen-states, which corresponds to the so called collective Lamb shift [9, 12-14].

The governing Hamilitonian for a quantum open system with a single continuum channel can be mapped to a plasmonic system consisting of an array of plasmonic dipole antennas with the same orientation positioned in proximity to each other (separations < wavelength). The mutual couplings among plasmonic resonators offer great flexibility in tailoring their optical properties, such as tuning the resonance frequency and controlling the line-shape of resonance [15-22]. Being an open system, the coupling can generally be divided into two parts, the direct near field coupling, and the indirect coupling mediated by radiative channels, i.e. continuum states. For antennas with small separations and the same orientation, the antennas primarily couple to the same dipolar radiation mode, i.e. a single decay channel with $K = 1$. In addition, the separations between the antennas can be carefully designed so that the effect of the direct near field coupling and the Cauchy principal value of the indirect coupling counteract each other, and the system is dominated by the anti-Hermitian part of the indirect coupling.

Recently, there have been explorations of electromagnetically induced absorption, or superscattering in plasmonic systems, which show optical effects opposite to the well-studied plasmon induced transparency [23, 24]. In those systems, the anti-Hermitian coupling plays an important role for introducing a constructive interference among different excitation pathways to enhance the scattering [25, 26]. In this work, we show that continuum mediated anti-Hermitian coupling in an array of densely positioned plasmonic antennas can be utilized for spatially manipulating light in the deep subwavelength scale. We consider an array of plasmonic antennas with different resonance frequencies $\omega_i \neq \omega_j$, and investigate the coupling phenomena of antennas under the excitation of a plane wave. The general form of the coupled equations can be written as,

$$\begin{bmatrix} -\omega+\omega_1-i\gamma_1 & \cdot & -\kappa_{1p} & \cdot & -\kappa_{1M} \\ \cdot & \cdot & \cdot & \cdot & \cdot \\ -\kappa_{p1} & \cdot & -\omega+\omega_p-i\gamma_p & \cdot & -\kappa_{pM} \\ \cdot & \cdot & \cdot & \cdot & \cdot \\ -\kappa_{M1} & \cdot & -\kappa_{Mp} & \cdot & -\omega+\omega_M-i\gamma_M \end{bmatrix} \begin{bmatrix} A_1 \\ \cdot \\ A_p \\ \cdot \\ A_M \end{bmatrix} = \begin{bmatrix} g_1 \\ \cdot \\ g_p \\ \cdot \\ g_M \end{bmatrix} E_0 \qquad (3)$$

where $\gamma_p$ is the dissipation term, $\kappa_{pq}$ ($\kappa_{pq} = \kappa_{qp}$) is the coupling coefficient between the *p*th and *q*th antennas, $g_p$ denotes the coupling strength between the incident plane wave and the antenna, with $g_p \propto V_p$, and $E_0$ is the electric field of the incident wave. For dipole antennas operating at infrared frequencies, the overall loss is usually dominated by the radiation loss (coupling to the continuum channel). For simplicity of analysis, we assume that $\gamma_p$ only contains the radiation loss. We further assume an ideal case that the Hermitian part of the overall coupling, which includes the direct near field coupling, and the Cauchy principal value of the indirect coupling, vanishes, and $\kappa_{pq}$ only consists of the indirect coupling mediated by the single decay channel. With the above assumptions, $i\gamma_p$ and $\kappa_{pq}$ are just the diagonal and off-diagonal elements of the anti-Hermitian coupling tensor $iVV^+/2$, respectively, i.e.

$$i\gamma_p = iV_p^2/2, \qquad \kappa_{pq} = iV_pV_q/2 \qquad (4)$$

At the resonance frequency of the *p*th antenna ($\omega = \omega_p$), the *p*th column of the coupling matrix in Eq. (3) is reduced to,

$$\begin{bmatrix} -\kappa_{1p} \\ \bullet \\ -i\gamma_p \\ \bullet \\ -\kappa_{Mp} \end{bmatrix} = -i\frac{V_p}{2}\begin{bmatrix} V_1 \\ \bullet \\ V_p \\ \bullet \\ V_M \end{bmatrix}$$

which has a linear relation with the excitation vector $[g_1, g_2 \ldots]^T$ on the right hand side of Eq. (3). Thus, Eq. (3) can be solved as,

$$A_p = \frac{2ig_p}{V_p^2}, \qquad A_q(q \neq p) = 0 \qquad (5)$$

Eq. (5) suggests that only a single resonator can be selectively excited at its resonance frequency, whereas all the others remain completely unexcited. In a strongly coupled system the eigen-states of the multi-antenna system are hybridized excitations of more than one resonator. This indicates that, under plane wave excitation of the antenna array, multiple eigen-states are simultaneously excited, which interfere constructively at a single resonator, and destructively at all the others, leading to a highly localized state. Since the spatial separation between the neighboring antennas can be much less than the wavelength of the incident electromagnetic wave, the plasmonic system governed by anti-Hermitian coupling offers an interesting platform for manipulation of light in the deep subwavelength scale.

The assumption of purely anti-Hermitian coupling constants among all the antennas is apparently too ideal in a realistic plasmonic system. Nonetheless, this condition can be relaxed, and only the coupling constants among those antennas with close resonance frequencies are critical for achieving a good contrast in selective excitation of individual antennas. To the first order approximation, a small deviation $\Delta\kappa_{pq}$ of the coupling coefficient $\kappa_{pq}$ from the ideal coupling strength of $-iV_pV_q/2$ results in a non-zero $A_q \sim \frac{\Delta\kappa}{\omega_q - \omega_p}A_p$ at the resonance frequency of the $p$th antenna. Therefore, the deviation of coupling constant from the idea case for two resonators with very different resonance frequencies has very little effect on the performance of the system.

A simple implementation of a plasmonic system exhibiting anti-Hermitian indirect coupling mediated by a single open channel is shown in Fig. 1. The system consists of two optical dipole antennas [27] of slightly different lengths with deep sub-wavelength spacing [Fig. 1(a)]. Each optical antenna is capable of focusing light into the gap region of nanometer scales. Numerical simulation (CST Microwave Studio) was carried out to calculate the optical response of each antenna at their gap centers under the illumination of a plane wave at normal incidence, and numerical fitting was subsequently employed to retrieve the coupling constant between the two antennas [28]. In the simulation, Drude model was used for the gold dielectric parameters, with $\omega_p = 1.37 \times 10^{16} \, rad/s$, and $\gamma = 4.1 \times 10^{13} \, rad/s$. The retrieved parameters are shown in Fig. 1 (b) for a range of separations $s$ between the two antennas.

The coupling constant has an imaginary part (red curve) that is only slightly less than the dissipation of both antennas (blue and green) over a broad range of $s$ up to about 250 nm. The approximate equity $\text{Im}(\kappa_{12})^2 \approx \gamma_1 \gamma_2$ indicates the presence of an anti-Hermitian term of rank 1 in the governing Hamiltonian of the plasmonic system, i.e. both antennas primarily couple to a single radiative decay channel. The slight difference between the anti-Hermitian coupling strength $\text{Im}(\kappa_{12})$ and the dissipations at very small antenna separation is mainly due to the intrinsic ohmic loss of the antennas. As $s$ increases above 250 nm, the effect of other radiative channels, such as electric quadrupole and magnetic dipole radiation modes, start to become significant, and the assumption of a single continuum channel does not hold anymore. Consequently, the imaginary part of the coupling constant deviates significantly from the dissipation rates of the antennas, and the foregoing analysis no longer applies in this regime.

At very small separations, the real (Hermitian) part of the coupling coefficient exhibits a large negative real part, which is dominated by the direct near-field coupling between the two antennas. With increasing distance, the near field coupling decreases rapidly, and the real part of the coupling crosses zero at a separation around 30 nm, where the direct near field coupling is cancelled by the Cauchy principal value of the indirect coupling, leaving a purely anti-Hermitian coupling between the two antennas. Around this separation, highly asymmetric Fano profiles for the response of each antenna are observed, as indicated by Fig. 1(c). There exists a strong suppression of the excitation of each antenna as manifested by the spectral dip at 225 and 241 THz, respectively (marked by the black and red dashed lines).

Although the spectral dips at these two frequencies do not go to zero as given by Eq. (5) for the ideal case, there appears a large contrast between the amplitudes of the two antennas. Thus, light localization can be switched from one antenna to the other by slightly tuning the frequency of incident light. As Fig. 1(b) shows, the imaginary part of coupling constant is greater than the real parts for a broad antenna separation range from 20 nm to 350 nm. Thus, the coupled optical antennas serve as a model system for studying the interesting physics in a physical open system dominated by the anti-Hermitian coupling.

Next we extend the plasmonic system to include five optical antennas with deep sub-wavelength spacing [Fig. 2(a)]. The length of the antennas ranges from 450 nm on one side to 610 nm on the other side with a step of 40 nm. The edge-to-edge spacing between the nearest neighboring antennas is 45 nm, and the separations between the second, third and forth nearest neighbors are 130 nm, 215 nm, 300 nm, respectively. As indicated by Fig. 1(b), this configuration gives rise to imaginary (anti-Hermitian) part dominated coupling coefficients among the antennas in the array. The anti-Hermitian part of the coupling constants roughly satisfy $\text{Im}(\kappa_{pq})^2 \approx \gamma_p \gamma_q$ up to the third nearest neighbor ($s = 215$ nm), but not for the two antennas at the two ends (forth nearest neighbor). However, as the two antennas at the two ends have dramatically different lengths and therefore very different resonance frequencies, this deviation has only a small effect on the overall performance. The major deviation from the ideal case described by Eqn (4) is that the there exists real (Hermitian) part of the coupling constants among the antennas. Nonetheless, it will be shown later through both numerical simulations and experimental observations that the selective excitation of individual antennas can still be achieved despite the presence of Hermitian coupling among the antennas.

We performed the numerical simulation on the antenna array using the commercial software, CST Microwave Studio$^{\text{TM}}$. In the simulation, a plane wave is incident at normal incidence onto the array of antennas. As shown in Fig. 2(b), each antenna exhibits a large resonance peak at approximately its uncoupled resonance frequency [Fig. 2(c)], where all the other antennas are strongly suppressed, leading to selective excitation of an individual antenna. This observation is consistent with the theoretical analysis given by Eq. (5), despite the presence of Hermitian coupling constants and ohmic loss that result in a deviation of $\kappa_{pq}$ from $i\gamma_p\gamma_q$ for interaction between certain antennas. In contrast to the selective excitation and sharp spectral features for the coupled antenna array, an array of uncoupled antenna exhibits very poor excitation selectivity due to the relative broad line-width of the resonance peak, as shown in Fig. 2(c). It is interesting to note from Fig. 2(b) that, besides the main resonance peak for each antenna, there exist several small spectral ripples with narrow line-widths. All these narrow spectral features are situated on top of a much broader spectral hump. This indicates the presence of several long-lived states super-imposed on a super radiant state with broad spectral line-width, which is characteristic of an open system. We note that similar selective excitation among a plasmonic antenna array has been theoretically proposed previously [29]. However, the fundamental mechanism of selective excitation is very different. In that work, the radiative loss and coupling constants among the antennas are both strongly suppressed by placing a ground plane underneath the antenna array. Whereas the coupling mediated by radiation plays a key role in achieving the selective excitation in the work presented here.

To visualize the selective excitation of individual antennas, we numerically obtained the out-of-plane electric field intensity distribution of the antenna array at the resonance frequencies of each antenna, as shown in Fig. 3(a-e). Note that the out-of-plane component of the electric field represents the charge distribution of the plasmonic system. The field distribution confirms that the antennas are individually excited at their resonance wavelengths, and the hot spots shifted from the shortest antenna at 1200 nm wavelength to the longest antenna at a wavelength of 1600 nm.

Experimentally, we fabricated the plasmonic antenna array and directly observed selective excitation of individual antennas at carefully chosen wavelengths, as shown in Fig. 3. The plasmonic antenna array was fabricated on a quartz wafer using electron beam lithography, followed by metal evaporation and a lift-off process. To characterize the optical response of the plasmonic antenna array, a scattering-type near-field optical scanning microscope (NSOM) was employed to measure the near-field distribution at the top surface of the plasmonic multiplexers. A supercontinuum light source together with an acoustic-optical modulator is used as a tunable light source in the near-infrared range with a linewidth of 5nm. Fig. 3(f-j) show the near field optical distribution for the array of optical antennas at five different wavelengths from 1200 nm to 1600 nm at a step of 100 nm. At each optical excitation wavelengths, only a single antenna was strongly excited, whereas all the others were suppressed, which is in good agreement with the simulation results. Note that the asymmetric field distribution in the measurement may be caused by the asymmetric shape of the NSOM tip because of the degradation of tip during the measurement. The NSOM measured localization of light in the antenna array is further quantified by a plot of intensity of light integrated along the $y$ direction, versus $x$, at each measured wavelength, as shown in Fig. 3(k). On average, light is strongly confined within a scale of 70 nm along the $x$ direction at each excited wavelength.

Both simulation and measurement show that light can be confined and manipulated in the deep subwavelength scale in an array of optical dipole antenna, as the center-to-center distance between the neighboring antennas is only 85 nm which is about 1/15 of the optical wavelengths. To highlight the role that the coupling among the antennas plays, a control sample which consists of an array of antennas with the same geometrical parameters, but spatially separated far (300 nm) from each other, was measured, at a wavelength of 1300 nm. At this separation, there exists a large difference between the imaginary part of the coupling constant and the dissipation rate of each antenna, and therefore a single open channel approximation does not apply. As shown in Fig. 4(l), there is no selective excitation of individual antennas as several antennas were simultaneously excited. The comparison between the control sample and the well-designed antenna array implies that the coupling among antennas mediated by a single decay channel plays a significant role in achieving selective excitation of individual antennas.

Harnessing the coupling constants among plasmonic elements is very important for achieving desirable optical functionalities. We show that for a plasmonic system consisting array of dipole antennas closely packed to each other, the coupling among them can be dominated by the imaginary part, therefore, it serves as an interesting classical analog to open quantum systems extensively studied in nuclear and atomic physics. While most studies on coupled plasmonic systems have focused on real coupling constants that primarily induce mode splitting [30-32], we show that coupling constants that are dominated by the imaginary part introduce the interesting functionality of light manipulation in the deep subwavelength scale. Nano-manipulation of optical hot spots has been realized by various means including time reversal, adaptively shaping the wave front and polarization of the incident beam [33-36], which, however, involves complicated temporal-spatial light modulation techniques. It has been shown numerically that a number of distinct patterns in an array of plasmonic particles could be excited by a plane wave; the method was based on a numerical searching process by varying all the possible incidence angles and wavelengths [37]. Whereas in our work, we have used a physics driven method, and show that non-Hermtian coupling mediated by a single open channel plays a significant role in selective excitation of a single element. It has also been experimentally demonstrated that in a linear array of plasmonic particles, the particles at the two ends can be selectively excited [38]. In contrast, we move a significant step further and experimentally demonstrate that each individual element can be selectively addressed, no matter the element is in the middle or at the ends of the array. The continuum mediated anti-Hermitian coupling in plasmonics brings new perspectives towards understanding and controlling light at deep subwavelength scale, offers a convenient way to address individual nanophotonic elements in a deep subwavelength nanoplasmonic circuits, and may find important applications in nanophotonic devices such as nanoscale demultiplexed photo-detectors.

Figure captions:

Fig. 1: Numerical retrieval of the indirect anti-Hermitian coupling between two optical antennas. (a) A plasmonic system consisting of two optical antennas made of gold on a quartz substrate. The lengths of the antennas are 430 nm and 470 nm. The width and thickness of both antennas are 40 nm and 25 nm, respectively. A beam is incident at normal direction onto the antennas, with polarization along the longitudinal direction of the antennas. (b) The retrieved real part (black) and imaginary part (red) of optical coupling constant between the two antennas, and the dissipations of the long (blue) and short antennas (green) as a function of their edge-to-edge separation $s$. (c) The simulated (solid) and fitted (dashed) spectra of the electric field probed at the center of the gap of two coupled antennas with lengths of 430 nm (black) and 470 nm (red) at separations of 30 nm, where the coupling between the two antennas is almost purely anti-Hermitian. The responses of the two antennas show highly asymmetric profiles, characteristic of Fano lineshape. The vertical dashed lines mark the eigen-frequencies in each figure.

Fig. 2: Numerical simulations on an array of antennas with anti-Hermitian coupling showing selective excitation of individual antennas by a plane wave. (a) The schematic of a plasmonic antenna array consisting of five optical antennas with gradually varying lengths, ranging from 430 nm to 590 nm at a step of 40 nm. The thickness and width of the metal strips are 25 nm and 40nm, respectively. The edge-to-edge separation between the nearest neighbors is 45 nm, which gives rise to a small real part of the coupling constant for the nearest neighbors. (b) The calculated electric field magnitude at the gap center of each antenna versus the optical wavelength. At the resonance of each antenna, all the other antennas are strongly suppressed, leaving a highly selective excitation of a single antenna.(c) The spectral response for five uncoupled antennas with the same geometries as (a).

Fig. 3: Experimental verification of the selective excitation of individual antennas in the plasmonic antenna array. (a-e) The simulated near-field distributions of the antenna array at five different wavelengths: 1200 nm (250 THz), 1300 nm (230.8 THz), 1400 nm (214.3 THz),

1500 nm (200 THz) and 1600 nm (187.5 THz), showing selective excitation of individual antennas. (f-j) The corresponding experimental observations show very good agreement with the simulations. The optical image (red), measured by NSOM, is overlapped onto the topographical image (grey) measured by the atomic force microscopy. (k) The measured intensity of light integrated along the *y* direction, versus *x*, at each measured wavelength. (l) The near field measurement at 1400 nm wavelength on the control sample, which consists of an array of antennas with the same geometry specification but at a large nearest neighbor separation of 300 nm. Without the anti-Hermitian coupling, all the antennas within the focus spot, which is around 2μm, are excited simultaneously due to the spectral overlapping of the resonances. Scale bar is 300nm.

Fig. 1

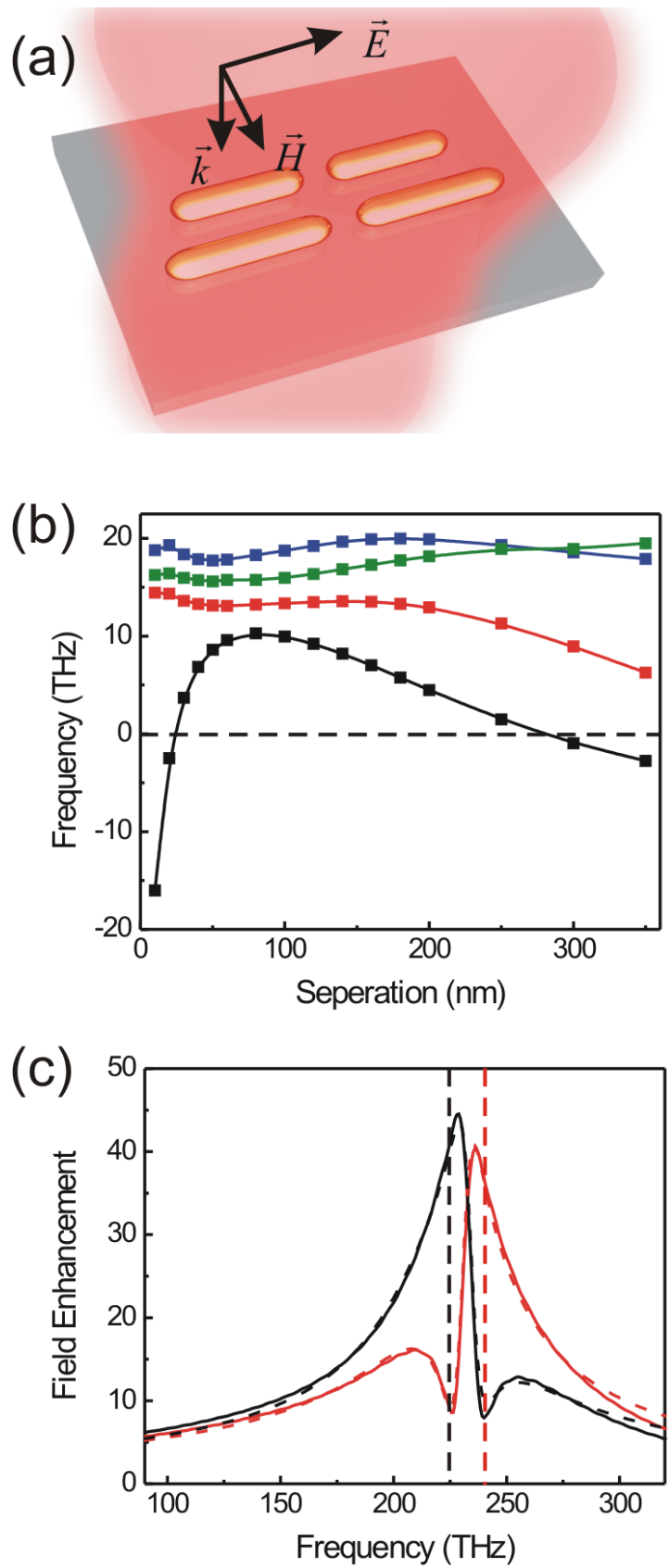

Fig. 2

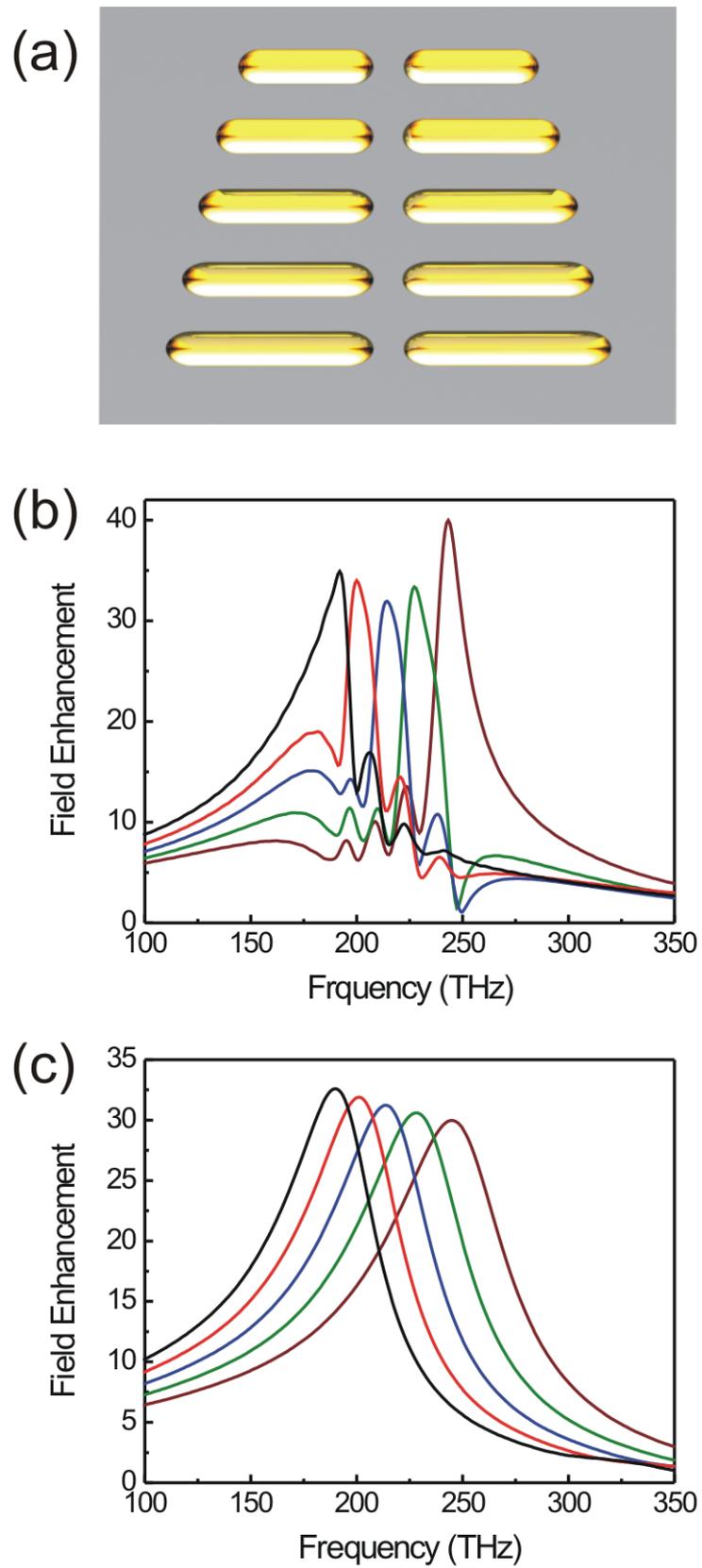

Fig. 3

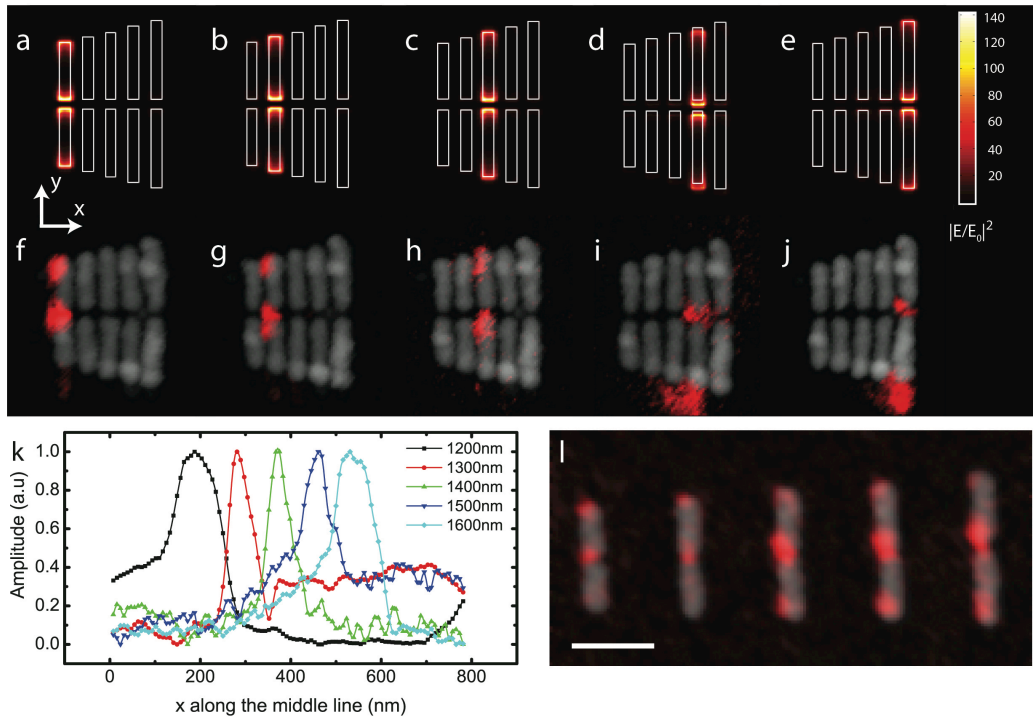